# Non-Hamilton cycle sets of having solutions and their properties

Heping Jiang [0000-0001-5589-808X]

jjhpjhp@gmail.com

## Abstract

A graph *G* is a tuple (*V*, *E*), where *V* is the vertex set, *E* is the edge set. A reduced graph is a graph of deleting non-Hamiltonian edges and smoothing out the redundant vertices of degree 2 on an edge except for leaving only one vertex of degree 2. A 2-common (*v*, *0*) combination is a cycle set in which every pair of joint cycles *A* and *B* satisfies |*V*(*A*)∩*V*(*B*)|=2 and |*E*(*A*)∩*E*(*B*)|=0. In this paper, we investigate the cycle structure of 2-common (*v*, *0*) combination in reduced graphs, and give the characterizations of their Hamiltoncity.

## 1   Introduction

In discrete mathematics and computer science, under what conditions that a graph is a Hamilton graph is a well known problem. There is a great amount of works on the problem, but no practical (or good) characterization for Hamilton graphs has been found [1].

Grinberg Theorem is one of few necessary conditions for Hamilton graphs, and has been regarding as a condition for planar Hamilton graphs. In [2], we gave a new combinatorial interpretation of Grinberg Theorem. Further investigation on the cycle combination showed that there only have two types of cycle combinations for the solution of a graph, one is that the sum of the cycles in the combination is a Hamilton cycle; while the other is not. The former is a 2-common (*v*, *e*) combination and the later a 2-common (*v*, *0*) combination.

In this paper, by examining the cycle structure in a 2-common (*v*, *0*) combination, we show that there is an area (marked as $C_k$) which looks like a part of the universal set. But from the point of view of a basis in the cycle space of a graph, this area should be a cycle so that the union of all the cycles generates the given graph. We characterize this area in detail and give three properties of the 2-common (*v*, *0*) combination (Lemma 3.1, 3.2, and 3.3).

## 2  Preliminaries

### Definitions

Graphs considered in this paper are finite, undirected, and simple connected graphs. A graph $G = (V, E)$ is a finite nonempty set *V* of elements called vertices, together with a set *E* of two elements subsets of *V* called edges. Let $B(G)$ be a basis of the cycle space of a graph *G*. Unless specified otherwise, any cycle in a basis is an elementary.

A vertex *v* is incident with an edge *e* if $v \in e$. The number of edges incident with a vertex is called the degree of the vertex. Two vertices *u*, *v* of *G* are neighbors if *uv* is an edge of *G*. Let $|P|$ be the number of the vertices of degree 2. We say a cycle is removable if removing the cycle from $B(G)$ produces a subgraph *G′* such that $V'=V$, $E'=E-1$ and the neighbors of any vertex in *G* satisfy $|P|<3$.

We denote by *R* the number of cycles passed through an edge in a graph *G*. A boundary vertex denotes a vertex that only has two edges of $R=1$ in its incident edges. A boundary edge refers to an edge of $R=1$. A cut vertex means a vertex that all its edges are $R=1$. A vertex is called inside if it is neither a boundary vertex nor a cut point.

In this paper, we use the equation for the equation (1.2) in [2]. The equation of a Hamilton graph has solutions. Sometimes we abbreviate such graph a solvable graph, in which there has a set of cycles in a basis satisfies the equation. We define the cycle set a solution set, where every cycle is called a solution cycle; and the complement of a solution set is called a co-solution set, where every cycle is called a co-solution cycle.

A cycle set $S$ is called a 2-common $(v, e)$ or a 2-common $(v, 0)$ combinatorial cycle set if every pair of joint cycles in $S$ is a 2-common $(v, e)$ or a 2-common $(v, 0)$ combination.

### 2-common $(v, 0)$ combination and $C_k$

In [2], we showed a new cycle set satisfied solvable graphs: a 2-common $(v, 0)$ combinatorial cycle set. Obviously, in a finite graph, we can combine $n$ cycles with two common vertices such that $|E(A) \cap E(B)|=0$, where $n \in \mathbb{Z}^+$, so that they generate a 2-common $(v, 0)$ combinatorial cycle set likes a graph in Figure 2.1. Since any pair of joint cycles satisfies the condition of $|V_i \cap V_j|=2$, then, without loss of generality, we can use one pair of joint cycles to represent a 2-common ($v$, $0$) combinatorial cycle set in our paper, unless stated specially. See the right graph in Figure 2.2.

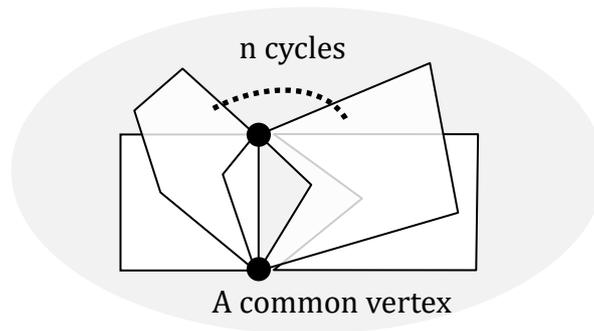

**Figure 2.1** Only for clearly displaying the combination, we use the grey color background to show a 2-common $(v, 0)$ cycle set with n cycles.

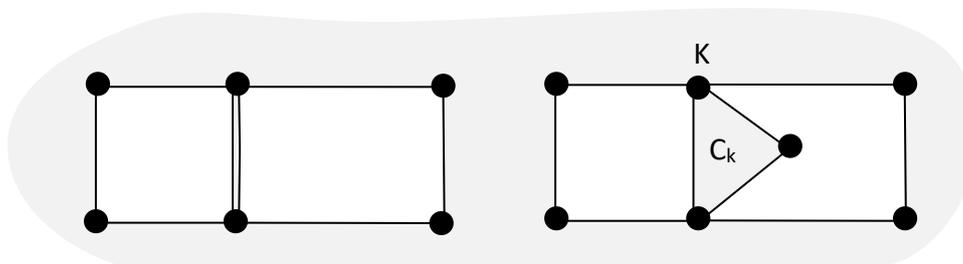

**Figure 2.2** The Left graph is a 2-common $(v, e)$ cycle set, the right graph is a 2-common $(v, 0)$ cycle set.

Let $G_{(v, 0)}$ be a graph that there is a 2-common ($v$,$0$) combinatorial cycle set in a basis of $G$. Let $K$ be a boundary vertex of degree 4. $C_k$ is a cycle on which all boundary vertices are $K$ but no edges are boundary, and $|C_k|$ is the number of $C_k$. A co-solution cycle in a graph $G$ is called unique if $G$ has a unique solution. Then, we give the following characterization of $C_k$ in $G_{(v, 0)}$,

**Proposition 2.1** *For a solvable graph $G_{(v, 0)}$, $C_k$ is an irremovable and unique co-solution cycle.*

*Proof.* First, we prove $C_k$ is not a removable cycle in $G_{(v, 0)}$, and second, $C_k$ is a unique co-solution cycle in $G_{(v, 0)}$.

By Lemma 1.2 in [2], a 2-common ($v$, $0$) combination is also a solution set to a graph $G$. Then, although the area $C_k$ (see Figure 2.2 right graph), by Venn Diagrams, is plainly a part of the universal set, it actually is a cycle in a basis of $G$. Clearly, by the definition, it is a co-solution cycle. Thus, the area $C_k$ should be a removable cycle. However, by the change of a basis, we can change the right graph in Figure 2.2 into the graph in Figure 2.3. Then, it is another basis of G, a 2-common ($v$, $e$) combination, where three combined

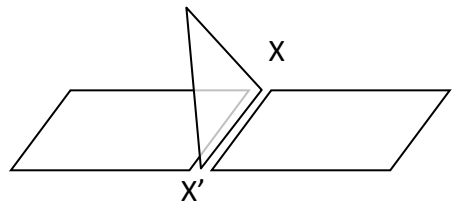

**Figure 2.3** A 2-common ($v$, $e$) cycle set

cycles share the edge XX′ such that every pair of two joint cycles is a

2-common ($v$, $e$) combination. Clearly, no matter on the vertex X or X′ there has $|P|≥3$. Note that a removable cycle satisfies $|P|<3$. Hence, $C_k$ is not a removable cycle although it should be removed from the basis of $G$.

$C_k$ in a 2-common ($v$, $0$) combination is a co-solution cycle, and by the definition, $C_k$ is the only cycle as a co-solution cycle because there have no other co-solution cycles to replace it in a basis of $G$. Hence, $C_k$ is clearly a unique co-solution cycle.∎

**Two basic rules and reduced graphs**

According to the definition of a Hamilton cycle, it is basically to determine some edges of no Hamilton cycle passing through. Jefferey A. Shufelt and Hans J. Berliner in [3] and Basil Vandegriend in [4] detailed and proved the rules about how to delete these edges from a given graph. In this paper, we use two of the rules and rewrite as the following.

**Rule 2.1** *If the neighbors of any vertex of G satisfy $|P|≥3$, then G is non-Hamiltonian.*

**Rule 2.2** *Every edge incident to a vertex of degree 2 is a Hamilton path.*

By Rule 2.1, if there is $|P|≥3$ in the neighbors of a vertex, then there have $|P-2|$ non-Hamiltonian edges, and then we can dismiss them in considering the Hamiltoncity of a graph, that means these edges can be deleted. By Rule 2.2, if there is $|P|≥2$ between vertex u and v, then the Hamiltoncity of any edge uv with $|P|≥2$ can be represented by the edge *uv* with $|P|=1$, that means only one vertex of degree 2 is enough for determining the Hamiltoncity of an edge. In this paper, we use a reduced graph for a graph produced from deleting non-Hamiltonian edges by Rule 2.1 and the normalizing by Rule 2.2. Graphs considering in the following content are reduced graphs, unless specified otherwise.

Definitions and terminologies not mentioned in this paper please refer to [5] and [6].

## 3 Characterizations of $C_k$

In this Section, we present some characterizations of $C_k$.

**Lemma 3.1** *For the neighbors of any vertex in a solvable graph G,* $|P| \geq 3 \Leftrightarrow |C_k| \neq 0$.

*Proof.* According to Lemma 1.2 in [2], we have to consider two types of cycle sets: a 2-common $(v, e)$ combination and a 2-common $(v, 0)$ combination. Since there have no vertices that their degree is equal to and more than 3 in 2-common $(v, e)$ combination, then there have no vertices K in the cycle set, we have $|C_k|=0$. Thus, for the neighbors of any vertex in $G$, we have $|P| \not\geq 3 \Rightarrow |C_k|=0$. For a 2-common $(v, 0)$ combination, by the characterizations of $C_k$ in $G_{(v, 0)}$ (Proposition 2.1), we know, for the neighbors of any vertex in $G$, that there have $|P| \geq 3$ and $C_k$, which implies $|P| \geq 3 \Rightarrow |C_k| \neq 0$. That completes the proof. ∎

**Lemma 3.2** *Let B (G) be a basis of a solvable graph G. For all the cycle $C \in B$ (G), if $C \neq C_k$, then G is a Hamilton graph.*

*Proof.* By Lemma 3.1, for every cycle $C \in B(G)$, $C \neq C_k$ means $|P| \not\geq 3$. Therefore, there has no 2-common $(v, 0)$ combination in $B(G)$. By Lemma 1.2 in [2], there must exist a 2-common $(v, e)$ combination for every pair of joint cycles in $B(G)$. By the definition, the sum of a 2-common $(v, e)$ combination is a Hamilton cycle. Hence, for every cycle $C \in B(G)$, if $C \neq C_k$, then $G$ is a Hamilton graph. ∎

**Lemma 3.3** *A solvable graph G is Hamiltonian $\Leftrightarrow |C_k|=0$.*

*Proof.* By Lemma 3.1, $|C_k| \neq 0 \Rightarrow |P| \geq 3$; and by Rule 2.1, $G$ is not a Hamilton graph. By the definition, for every cycle $C \in B(G)$, $C \neq C_k$ can be expressed by $|C_k|=0$; and by Lemma 3.2, we know $G$ is a Hamilton graph. Thus, we complete the proof. ∎